\documentclass{article}
\usepackage{spconf,amsmath,graphicx}
\usepackage{multirow}
\usepackage{pifont}
\usepackage{balance}
\usepackage[numbers,sort&compress]{natbib}
\usepackage[colorlinks,linkcolor=blue]{hyperref}

\title{Residual Dense Swin Transformer for Continuous Depth-Independent Ultrasound Imaging}
\name{Jintong Hu, Hui Che, Zishuo Li, Wenming Yang$^{\dagger}$\thanks{This work was partly supported by the National Natural Science Foundation of China (Nos.62171251\&62311530100) and the Special Foundations for the Development of Strategic Emerging Industries of Shenzhen 
(Nos.JSGG20211108092812020\&CJGJZD20210408092804011).}}
\address{Tsinghua Shenzhen International Graduate School, Tsinghua University}
\begin{document}
%
\maketitle
\begin{abstract}
Ultrasound imaging is crucial for evaluating organ morphology and function, yet depth adjustment can degrade image quality and field-of-view, presenting a depth-dependent dilemma. Traditional interpolation-based zoom-in techniques often sacrifice detail and introduce artifacts. Motivated by the potential of arbitrary-scale super-resolution to naturally address these inherent challenges, we present the Residual Dense Swin Transformer Network (RDSTN), designed to capture the non-local characteristics and long-range dependencies intrinsic to ultrasound images. It comprises a linear embedding module for feature enhancement, an encoder with shifted-window attention for modeling non-locality, and an MLP decoder for continuous detail reconstruction. This strategy streamlines balancing image quality and field-of-view, which offers superior textures over traditional methods. Experimentally, RDSTN outperforms existing approaches while requiring fewer parameters. In conclusion, RDSTN shows promising potential for ultrasound image enhancement by overcoming the limitations of conventional interpolation-based methods and achieving depth-independent imaging.
\end{abstract}
\section{Introduction}
\label{sec:intro}

Ultrasound imaging serves as a pivotal tool in medical diagnostics for its non-invasive nature and real-time imaging capabilities, allowing visualization of superficial and deep structures \cite{10.3389/fphy.2020.00124}. However, adjusting the imaging depth presents challenges that impact image quality and field-of-view \cite{DepthEffect}.

\begin{figure}[htb]
\begin{minipage}[b]{1.0\linewidth}
  \centering
  \centerline{\includegraphics[width=8.5cm]{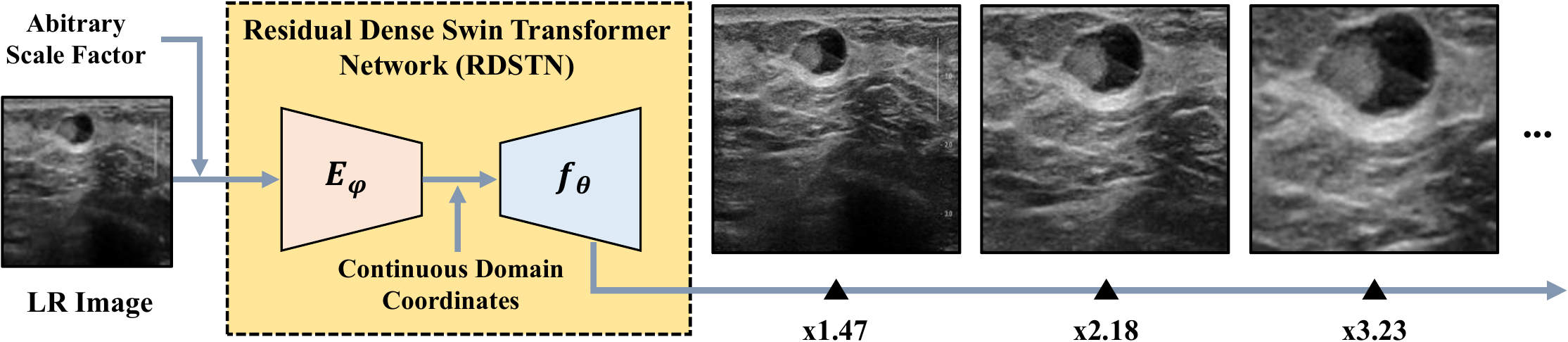}}
\end{minipage}
\caption{An example of magnified images of abitrary scales generated by RDSTN.}
\vspace{-0.2cm}
\label{fig:1}
\end{figure}

Modifying the imaging depth in ultrasound requires altering the echo reception time. Longer reception times, necessary for deeper imaging, tend to lower the frame rate, subsequently reducing temporal resolution \cite{GUNABUSHANAM2021100766}. A shallow imaging depth, however, may lead to interference from adjacent echo signals, compromising image quality. Therefore, selecting the appropriate depth threshold is crucial.

Traditionally, zoom-in operations utilizing interpolation have been employed to counterbalance unsatisfactory image quality during depth adjustments \cite{zoom}. This often results in the loss of intricate details and the emergence of aliasing artifacts.

Addressing this challenge, our study presents the arbitrary-scale super-resolution (ASSR) as a cutting-edge approach that offers an effective solution within the desired depth threshold. Existing ASSR models often overlook the unique attributes of datasets, a particularly critical aspect given the notable long-range similarities in ultrasound images.

To address the shortcoming, we introduce the Residual Dense Swin Transformer Network (RDSTN), which integrates a linear embedding layer, a Residual Dense Shifted-window Transformer (RDST) encoder, and a Multilayer Perceptron (MLP) decoder. The linear embedding layer projects the input into higher-dimensional space for enhanced feature extraction. The RDST encoder leverages non-locality to promote essential feature reuse. Finally, the MLP decoder maps each target pixel coordinate to its corresponding latent code, achieving refined reconstruction of detail. Figure \ref{fig:1} shows an example of magnified images generated by RDSTN.

\begin{figure*}
  \centering
  \includegraphics[width=\textwidth]{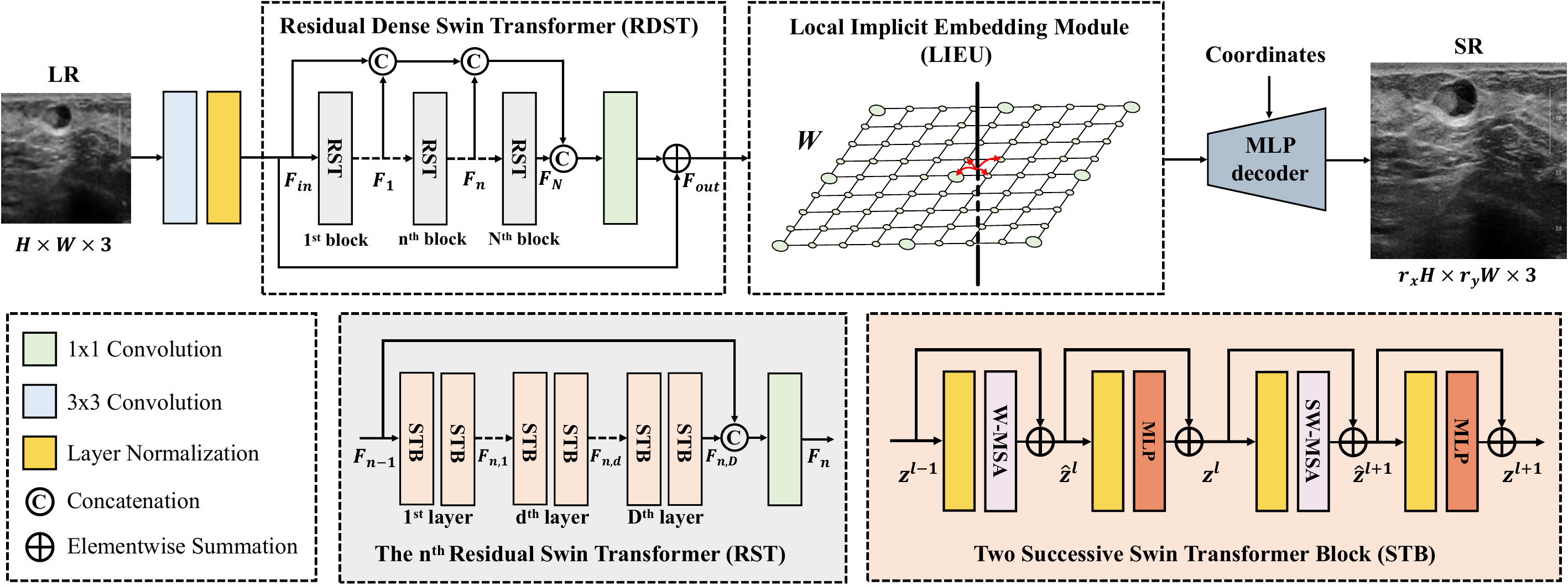}
  \caption{\textbf{The main pipeline of our RDSTN.} RDSTN introduces non-locality and allows for essential feature reuse, improving representation and performance.}
  \label{fig:2}
\end{figure*}

In conclusion, our contributions are:
\begin{itemize}
    \item The first introduction of implicit neural representation-based arbitrary-scale super-resolution to multiple organ ultrasound datasets, ensuring a versatile and depth-independent display.
    \item Proposing an efficient non-local encoder to model the spatial dependencies between pixels, reducing parameters by 45
    $\%$ with respect to the state-of-the-art method.
    \item Conducting extensive experiments to prove the capability of RDSTN in addressing the balance issue.
\end{itemize}

\section{Related Methods}
\label{sec:related}

\subsection{Single Image Super-Resolution}

Numerous single image super-resolution (SISR) methods have been proposed over time. Among the earlier techniques, interpolation stands out, which works by filling values between pixels to enhance image resolution. However, its major downside is the inability to restore intricate details effectively. Recently, deep learning-based methods have taken the spotlight in the SISR domain. Notably, Convolutional Neural Networks (CNNs) are now widely recognized due to their prowess in autonomously learning hierarchical features and establishing a relationship between low-resolution (LR) and high-resolution (HR) images. The SRCNN \cite{dong2015image}, a notable CNN-based model, leverages a three-layer CNN structure to learn this LR to HR mapping. Various other models, such as VDSR \cite{kim2016accurate}, EDSR \cite{lim2017enhanced}, ESPCN \cite{shi2016realtime}, LapSRN \cite{lai2017deep}, MemNet \cite{tai2017memnet}, RDN \cite{zhang2018residual}, and DBPN \cite{haris2018deep}, have carved their niche by adopting unique strategies. These include deep architectures, sub-pixel convolutional layers, the Laplacian pyramid framework, memory blocks, cascaded ResNet-like blocks, densely connected residual blocks, and deep back-projection units. However, a common limitation across these models is their lack of flexibility due to predefined upsampling mechanisms, thereby confining their application to fixed integer-scale super-resolution.

\subsection{Arbitrary-Scale Image Super-Resolution}

Arbitrary-scale super-resolution (ASSR) techniques are gaining traction due to their adaptability in handling super-resolution tasks across various scaling factors \cite{8578168,9008787,hu2019metasr,chen2021learning,lee2022local,wu2021scaleaware,li2022adaptive}. MetaSR \cite{hu2019metasr} led the charge in CNN-based methods for ASSR, while LIIF \cite{chen2021learning} showcased an innovative framework by leveraging implicit neural representations to treat images as continuous functions. Specifically, LIIF uses a Multi-Layer Perceptron (MLP) to determine RGB values from feature maps, coordinates, and scaling factors. Building on the foundation set by LIIF, other ASSR studies have emerged. For instance, LTE \cite{lee2022local} inputs coordinates into a high-dimensional Fourier feature space to tackle spectral bias. SADN \cite{wu2021scaleaware} augments the latent code drawn by the encoder with multi-scale features. A-LIIF \cite{li2022adaptive} addresses artifacts by capturing pixel differences using multiple MLPs. Nevertheless, a limitation of these methods is their dependence on convolution or fully connected layers, which curtails the encoder's capacity to glean global information and might bypass crucial spatial similarities among pixels. Consequently, integrating non-locality into the encoder is imperative.

\begin{table*}[!h]
    \centering
    \setlength{\abovecaptionskip}{0pt}
    \setlength{\belowcaptionskip}{8pt}
    \footnotesize
    \setlength{\tabcolsep}{9pt}
    \renewcommand{\arraystretch}{1.1}
    \caption{\textbf{Quantitative comparison in terms of PSNR(dB).} The evaluation is performed on the BUSI testing set. The models are trained with continuous scale sampled from $U(1, 4)$. Best result of each scale is in bold.}
    \label{table:1}
    \vspace{6pt}
    \begin{tabular}{c|c|c c c c c c c|c c c}
    \hline
        \multirow{2}{*}{Methods} & Num. of & \multicolumn{7}{c|}{In-distribution} & \multicolumn{3}{c}{Out-distribution} \\ \cline{3-12}
        ~ & Parameters & ×1.6 & ×1.7 & ×1.8 & ×1.9 & ×2 & ×3 & ×4 & ×6 & ×8 & ×10  \\ \hline
        Bicubic & -- & 40.21 & 39.36 & 38.88 & 38.21 & 38.68 & 33.17 & 30.40 & 26.88 & 24.86 & 23.64 \\ 
        EDSR-LIIF \cite{chen2021learning} & 496.4K & 43.92 & 43.06 & 42.26 & 41.50 & 40.80 & 35.80 & 32.87 & 29.42 & 27.34 & 26.04 \\
        RDN-LIIF \cite{chen2021learning} & 5.8M & 44.71 & 43.81 & 43.03 & 42.28 & 41.57 & \textbf{36.36} & \textbf{33.22} & 29.58 & 27.46 & 26.12 \\ 
        Unet \cite{ronneberger2015unet} & 31.4M & 42.39 & 41.71 & 41.05 & 40.42 & 39.83 & 35.24 & 32.55 & 29.20 & 27.16 & 25.91 \\ 
        Resnet50 \cite{he2015deep} & 4.1M & 42.86 & 42.07 & 41.35 & 40.62 & 39.95 & 35.17 & 32.46 & 29.12 & 27.11 & 25.87 \\ 
        RDSTN (ours) & 3.2M & \textbf{44.78} & \textbf{43.89} & \textbf{43.10} & \textbf{42.35} & \textbf{41.62} & 36.34 & 33.20 & \textbf{29.64} & \textbf{27.54} & \textbf{26.18} \\ \hline
    \end{tabular}
\end{table*}
\vspace{-0.3cm}

\begin{table*}[!h] 
    \centering
    \setlength{\abovecaptionskip}{0pt}
    \setlength{\belowcaptionskip}{8pt}
    \footnotesize
    \setlength{\tabcolsep}{9pt}
    \renewcommand{\arraystretch}{1.1}
    \caption{\textbf{Ablation study of RDSTN on Local Feature Fusion (LFF) and Global Feature Fusion (GFF).} The evaluation is performed on the BUSI testing set, with a focus on measuring the peak signal-to-noise ratio (PSNR(dB)) to assess the performance of these strategies. The best result of each scale is in bold.}
    \label{Ablation study}
    \vspace{6pt}
    \begin{tabular}{c|c|c|c c c c c c c|c c c}
    \hline
        \multirow{2}{*}{Model Settings} & \multicolumn{2}{c}{Module} & \multicolumn{7}{|c|}{In-distribution} & \multicolumn{3}{c}{Out-distribution} \\ \cline{2-13}
        ~ & LFF & GFF & ×1.2 & ×1.4 & ×1.6 & ×1.8 & ×2 & ×3 & ×4 & ×6 & ×8 & ×10  \\ \hline
        $S_{1}$ & \ding{56} & \ding{56} & 48.65  & 46.17  & 44.36  & 42.69  & 41.21  & 36.04  & 33.01  & 29.51  & 27.42  & 26.07   \\ 
        $S_{2}$ & \ding{56} & \ding{52} & 48.71  & 46.23  & 44.42  & 42.73  & 41.27  & 36.07  & 33.03  & 29.54  & 27.46  & 26.11   \\
        $S_{3}$ & \ding{52} & \ding{56} & 48.89  & 46.40  & 44.61  & 42.94  & 41.46  & 36.23  & 33.13  & 29.59  & 27.50  & 26.14   \\ 
        $S_{4}$ & \ding{52} & \ding{52} & \textbf{49.27}  & \textbf{46.62}  & \textbf{44.78}  & \textbf{43.10}  & \textbf{41.62}  & \textbf{36.34}  & \textbf{33.20}  & \textbf{29.64}  & \textbf{27.54}  & \textbf{26.18}   \\ 
        \hline
    \end{tabular}
\end{table*}

\section{Model Architecture}
\label{sec:model}

This section introduces the proposed Residual dense swin transformer network (RDSTN). The main pipeline of our RDSTN is shown in Figure \ref{fig:2}.

\subsection{Residual Dense Swin Transformer Encoder}

Swin Transformer block \cite{liu2021swin} has emerged as a highly promising architecture due to its unique capabilities in handling computation cost and non-local characteristics. Our proposed  Residual Dense Swin Transformer (RDST) encoder is designed on the basis of swin transformer block, composed of four hierarchical stages, each with six consecutive swin transformer blocks. The output of each stage maintains a fixed resolution and number of channels, thus the latent codes extracted by RDST have the same spatial resolution as the LR counterparts. Therefore, every pixel in low-resolution images can match a latent code in the same position. RDST assists the model to extract non-locality in broader receptive fields at the beginning stages, while retaining fine-grained detail at the higher stages. 

A distinctive innovation in RDST is the fusion of both local and global features, enabling the model to retain vital contextual information throughout its processing stages. This is achieved by adeptly concatenating outputs from each stage, thereby harnessing the rich feature spectrum across varying layers and blocks. To foster rapid model convergence, we incorporate residual connections with the originating image. Additionally, RDST adopts a cyclic interplay among successive layers, a strategic maneuver ensuring that the all-encompassing non-local information is captured with refined precision.

Specifically, global feature fusion and local feature fusion can be formulated as:
\begin{equation}
F_{out}=F_{in}+Conv_{1 \times 1}([F_{in}, F_{1}, \cdots, F_{N}])
\end{equation}
\begin{equation}
F_{n}=Conv_{1 \times 1}([F_{n-1}, STB^{D}(F_{n-1})])
\end{equation}
where STB denotes the swin transformer blocks. $F_{i}$ denotes the output of the $i^{th}$ RST block $(i\in{1, \cdots, N})$. $Conv_{1 \times 1}$ represents the convolution operation with 1×1 kernel size, which is used to recover channels.

\subsection{Local Enhanced Implicit Representation Upscaling}

Building on the paradigm of LIIF \cite{chen2021learning}, we introduce the Local Enhanced Implicit Representation Upscaling (LEIRU) decoder. This refined Multi-Layer Perceptron (MLP) is tailored to operate on the coordinates of the anticipated scale super-resolution image, aligning each coordinate with its nearest latent code. For a singular coordinate, denoted as $x_{q}$, its corresponding nearest latent code is represented as $C(x_q)$. The RGB values for every coordinate are thus defined by:
\begin{equation}
RGB(x_{q})=MLP([C(x_q), x_q-x^{*}])
\end{equation}
where MLP represents an MLP with learnable parameters, $x^{*}$ denotes the coordinate of $C(x_q)$. 

In essence, the relative distance gauges the affinity between features and latent codes. A closer coordinate to the latent code implies higher similarity, gauged by the inverse of this relative distance. The underlying premise is that coordinates within the same grid derive from the same latent code, thereby utilizing local information. The non-local encoder's design is pivotal to infuse non-locality into the local decoder, optimizing the model's performance.

However, an issue arises when a coordinate traverses the grid boundary. At this juncture, the MLP output can shift dramatically due to the abrupt latent code alteration, resulting in chessboard artifacts in the output image. To mitigate this, our module introduces a local ensemble operation. Drawing from interpolation principles, the local ensemble harnesses the weights of the rectangle, formed by the coordinate and its four adjacent latent codes, executing a nearest neighbor interpolation. Mathematically, this is articulated as:
\begin{equation}
LEIRU(x_{q})=\sum_{x_{i}\in{grid}} w_{x_{i}}RGB(x_{i})
\end{equation}
where $w_{x_{i}}$ represents the weights for each coordinate $x_{i}$. $RGB(x_{i})$ represents the corresponding RGB values computed by the MLP for the coordinate $x_{i}$.

\begin{figure*}
  \centering
  \includegraphics[width=\textwidth]{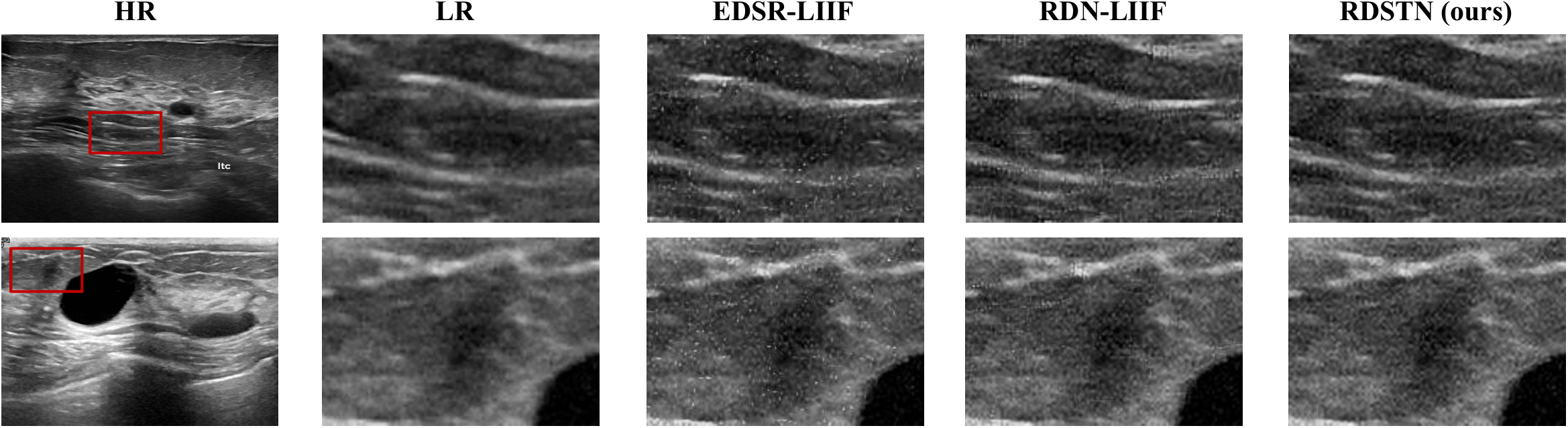}
  \vspace{-0.6cm}
  \caption{\textbf{Visual comparison between methods.} RDSTN achieves better continuity in texture and has the fewest white noise points compared to other methods.}
  \label{fig:3}
\end{figure*}

\section{Experiment Results}
\label{sec:result}

We utilize the BUSI dataset \cite{ALDHABYANI2020104863}, which comprises 779 high-resolution breast ultrasound images, to train our model. We allocate 80$\%$ of these images for training, with the remainder designated for testing. In line with prior research, our primary evaluation metric is the peak signal-to-noise ratio (PSNR).

\subsection{Comparison with other methods}

We benchmark our RDSTN against other implicit representation techniques, such as EDSR-LIIF, RDN-LIIF \cite{chen2021learning}, and the conventional bicubic interpolation \cite{1163711}. Table \ref{table:1} showcases the qualitative results. Across various scales, both within and beyond distribution, our method surpasses contemporary techniques while utilizing fewer parameters. When tested on low-resolution images tainted with Gaussian noise, the visual evidence in Figure \ref{fig:3} reveals our method's superiority in minimizing white noise spots and maintaining texture continuity. Therefore, our approach not only excels in standard metrics but also demonstrates impressive noise resistance.

\subsection{Ablation Study}

To further validate the efficacy of our model, we perform an ablation study on global and local feature fusion strategies. Our experiments indicate that setting $S_{4}$ yields the best results among all the tested scales, followed by $S_{3}$, $S_{2}$, and $S_{1}$, respectively. This underscores the effectiveness of our designed local and global feature fusion, with detailed results provided in Table \ref{Ablation study}. By integrating both global and local feature fusion, our network enhances its ability to capture non-local features while simultaneously reducing computational costs.

\begin{table}[!ht]
    \centering
    \footnotesize
    \setlength{\tabcolsep}{9pt}
    \caption{\textbf{Generalization test of RDSTN on out-distribution dataset.} The evaluation is performed on the 2023 MICCAI ultrasound enhancement challenge dataset, we use PSNR(dB) to assess the performance. The best result is in bold.}
    \label{table:3}
    \vspace{6pt}
    \renewcommand{\arraystretch}{1.1}
    \begin{tabular}{c|c c c c c}
    \hline
        \multirow{2}{*}{Method} & \multicolumn{5}{c}{scale} \\ \cline{2-6}
        ~ & ×1.6 & ×1.7 & ×1.8 & ×1.9 & ×2  \\ \hline
        \multicolumn{6}{c}{train: BUSI \cite{ALDHABYANI2020104863}, test: MICCAI USenhance breast} \\ \hline
        Bicubic & 34.28  & 33.52  & 31.55  & 31.23  & 31.66   \\ 
        EDSR-LIIF & 35.30  & 34.55  & 33.63  & 33.07  & 32.42   \\ 
        RDN-LIIF & 35.12  & 34.41  & 33.56  & 33.13  & 32.47   \\
        RDSTN (ours) & \textbf{35.35} & \textbf{34.62} & \textbf{33.74} & \textbf{33.23} & \textbf{32.59} \\ \hline
        \multicolumn{6}{c}{train: BUSI \cite{ALDHABYANI2020104863}, test: MICCAI USenhance thyroid} \\ \hline
        Bicubic & 38.17  & 37.14  & 34.91  & 34.26  & 34.81   \\ 
        EDSR-LIIF & 39.87  & 38.73  & 37.72  & 36.86  & 36.07   \\ 
        RDN-LIIF & 39.91  & 38.77  & 37.76  & 36.90  & 36.10   \\ 
        RDSTN (ours) & \textbf{39.99} & \textbf{38.81} & \textbf{37.83} & \textbf{36.96} & \textbf{36.14} \\ \hline
        \multicolumn{6}{c}{train: BUSI \cite{ALDHABYANI2020104863}, test: MICCAI USenhance carotid} \\ \hline
        Bicubic & 38.11  & 37.14  & 34.95  & 34.37  & 34.84   \\ 
        EDSR-LIIF & 40.05  & 38.96  & 37.90  & 37.07  & 36.21   \\ 
        RDN-LIIF & 40.08  & 38.98  & 37.89  & 37.11  & 36.28   \\ 
        RDSTN (ours) & \textbf{40.16} & \textbf{39.07} & \textbf{37.99} & \textbf{37.17} & \textbf{36.30} \\ \hline
    \end{tabular}
    \vspace{-0.4cm}
\end{table}

\subsection{Generalization}

To further assess the generalization capability of our model, we train it on breast ultrasound images and subsequently test on different tissues, including thyroid and carotid arteries. The results, present in Table \ref{table:3}, show that our model surpasses other methods, highlighting its strong generalization potential. Thus, our model can deliver exceptional results on expansive samples even when trained on a small, focused dataset, underscoring its adaptability in practical applications.

\section{Conclusion}

Our advanced RDSTN effectively tackles the challenges associated with long-range modeling and non-local feature extraction in arbitrary-scale super-resolution. It streamlines the delicate balance between image quality and field-of-view, showcasing enhanced noise suppression capabilities. Testing reveals that RDSTN performs competitively in both metrics and visual quality compared to other methods, yet utilizes fewer parameters. Through RDSTN, we can adeptly navigate continuous imaging at suitable depth thresholds. 

\section{Acknowledgments}
\label{sec:acknowledgments}

We wish to extend our heartfelt appreciation to the organizers of the 2023 Ultrasound Image Enhancement (USenhance) Challenge for availing the dataset pivotal for our generalization examination. Dataset can be accessed via the official 2023 USenhance portal: https://ultrasoundenhance2023.grand-challenge.org/.

\bibliographystyle{IEEEbib}
\bibliography{paper}

\end{document}